\documentclass[10pt,twocolumn,letterpaper]{article}

\usepackage{cvpr}              
\usepackage[accsupp]{axessibility}

\usepackage[dvipsnames]{xcolor}

\definecolor{cvprblue}{rgb}{0.21,0.49,0.74}
\usepackage[pagebackref,breaklinks,colorlinks,citecolor=cvprblue]{hyperref}
\usepackage{algorithm}
\usepackage{algpseudocode}
\usepackage{multirow}

\usepackage{colortbl}
\usepackage{array}
\usepackage[utf8]{inputenc}
\usepackage{mathtools}

\usepackage{subcaption}
\usepackage{booktabs} 

\def\paperID{14343} 
\def\confName{CVPR}
\def\confYear{2024}

\title{Communication-Efficient Collaborative Perception via \\ Information Filling with Codebook}

\author{Yue~Hu$^{1,4}$ \quad Juntong~Peng$^{1,4}$ \quad Sifei~Liu$^{1,4}$ \quad Junhao~Ge$^{1,4}$\quad Si~Liu$^{3}$ \quad Siheng Chen$^{1,2,4}$\\
$^{1}${Cooperative Medianet Innovation Center, Shanghai Jiao Tong University} \quad $^{2}${Shanghai AI Laboratory} \\ \quad $^{3}${Beihang University}  \quad 
$^{4}${Multi-Agent Governance \& Intelligence Crew (MAGIC)}\\ 
$^{1}${\tt\small \{18671129361,juntong.peng,hiraeth416,cancaries,sihengc\}@sjtu.edu.cn}  \quad 
$^{3}${\tt\small \{liusi\}@buaa.edu.cn} \\
}

\begin{document}
\maketitle


\begin{abstract}

Collaborative perception empowers each agent to improve its perceptual ability through the exchange of perceptual messages with other agents. It inherently results in a fundamental trade-off between perception ability and communication cost. To address this bottleneck issue, our core idea is to optimize the collaborative messages from two key aspects: representation and selection. The proposed codebook-based message representation enables the transmission of integer codes, rather than high-dimensional feature maps. The proposed information-filling-driven message selection optimizes local messages to collectively fill each agent's information demand, preventing information overflow among multiple agents. By integrating these two designs, we propose~\texttt{CodeFilling}, a novel communication-efficient collaborative perception system, which significantly advances the perception-communication trade-off and is inclusive to both homogeneous and heterogeneous collaboration settings. We evaluate~\texttt{CodeFilling} in both a real-world dataset, DAIR-V2X, and a new simulation dataset, OPV2VH+. Results show that~\texttt{CodeFilling} outperforms previous SOTA Where2comm on DAIR-V2X/OPV2VH+ with 1,333/1,206$\times$ lower communication volume. Our code is available at~\url{https://github.com/PhyllisH/CodeFilling}.
\end{abstract}
\section{Introduction}
\label{sec:intro}

\begin{figure}[!t]
    \centering
    \includegraphics[width=1.0\linewidth]{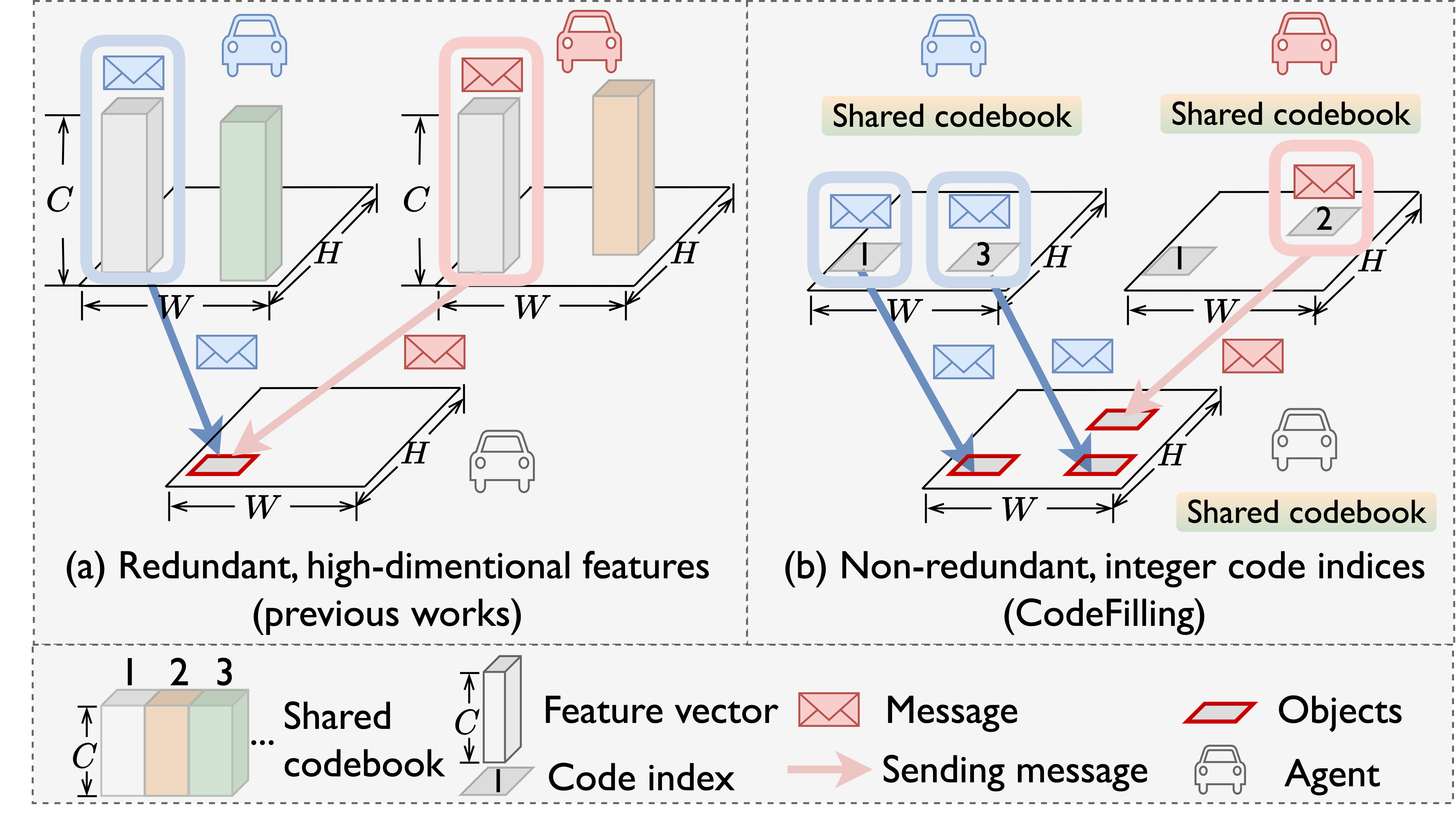}
  \vspace{-7mm}
  \caption{\texttt{CodeFilling} avoids redundant messages and achieves more complete detections by transmitting more critical perceptual information with the compact code index message.}
  \vspace{-7mm}
  \label{fig:intro}
\end{figure}

Collaborative perception aims to enhance the perceptual ability of each individual agent by facilitating the exchange of complementary perceptual information among multiple agents~\cite{WangV2vnet:ECCV20,YuDAIRV2X:CVPR22,LiuWhen2com:CVPR20,HuWhere2comm:NeurIPS22,LiLearning:NeurIPS21,HuCollaboration:CVPR23,SchererAutonomous:MAV15,AlotaibiLsar:IA19,LiS2RViT:ArXiv23,CoBEVFlow:NeurIPS23,HEAL:ICLR24}. It fundamentally overcomes the occlusion and long-range issues in single-agent perception~\cite{LangPointPillars:CVPR2018,ZhouObjects:Arxiv2019,Chen3DPC:SPM21,HuAerial:RAL22}.
As the forefront of autonomous systems, collaborative perception shows significant potential in enormous real-world applications, particularly vehicle-to-everything-communication-aided autonomous driving~\cite{WangV2vnet:ECCV20,WangCORE:ICCV23,ChenFcooper:SEC19,WangUMCAU:CVPR2023,LiAmongUS:ICCV23,XiangHMViT:ICCV23,XuV2V4Real:CVPR23,ChenCO3:ICLR23,SuUncertainty:ICRA23,XuBridging:ICRA23}.

In this emerging field, a central challenge lies in optimizing the trade-off between perception performance and communication cost inherent in agents sharing perceptual data~\cite{LiuWhen2com:CVPR20,LiuWho2com:ICRA20,HuWhere2comm:NeurIPS22,HuCollaboration:CVPR23,WangV2vnet:ECCV20,LiLearning:NeurIPS21,YangSpatioTemporalDA:ICCV2023,YangWhat2comm:ICCV2023}. Given inevitable practical constraints of communication systems, efficient utilization of communication resources is the prerequisite for collaborative perception. To minimize communication overhead, a straightforward solution is late collaboration, where agents directly exchange the perception outputs. However, numerous previous works indicate that late collaboration yields marginal perception improvements and is vulnerable to various noises~\cite{LuRobust:ICRA23,HuWhere2comm:NeurIPS22}. To optimize the perception-communication trade-off, most studies consider intermediate collaboration, where the collaborative messages are perceptual features~\cite{XuOPV2V:ICRA22,XuV2XViT:ECCV22,LiLearning:NeurIPS21,LiuWhen2com:CVPR20,LiuWho2com:ICRA20,HuWhere2comm:NeurIPS22,HuCollaboration:CVPR23,How2comm:NeurIPS2023,WangUMCAU:CVPR2023}. For example, When2com~\cite{LiuWhen2com:CVPR20} proposed the handshake strategy to limit the number of collaborators; and Where2comm~\cite{HuWhere2comm:NeurIPS22} proposed a pragmatic strategy that only transmits messages about constrained spatial areas. While these methods mitigate certain communication costs, they still necessitate the transmission of high-dimensional feature maps, which incurs substantial communication expenses.

To overcome the limitations of intermediate collaboration, our core idea is to optimize the collaborative messages from two key perspectives: representation and selection. For message representation, we introduce a codebook to standardize the communication among agents, where each code is analogous to a word in the human language dictionary. Based on this shared codebook among all agents, we can use the codes to approximate perceptual features; consequently, only integer code indices need to be exchanged, eliminating the need for transmitting high-dimensional features comprised of floating-point numbers. For message selection, we propose an information filling strategy, akin to piecing together a jigsaw puzzle. In this approach, assuming an agent's information demand is upper bounded, each of its collaborators performs a local optimization to select non-redundant messages to fill its information gap. This strategy prevents information overflow among multiple agents, further significantly reducing communication cost.

\begin{figure*}[!t]
    \centering
    \includegraphics[width=1.0\linewidth]{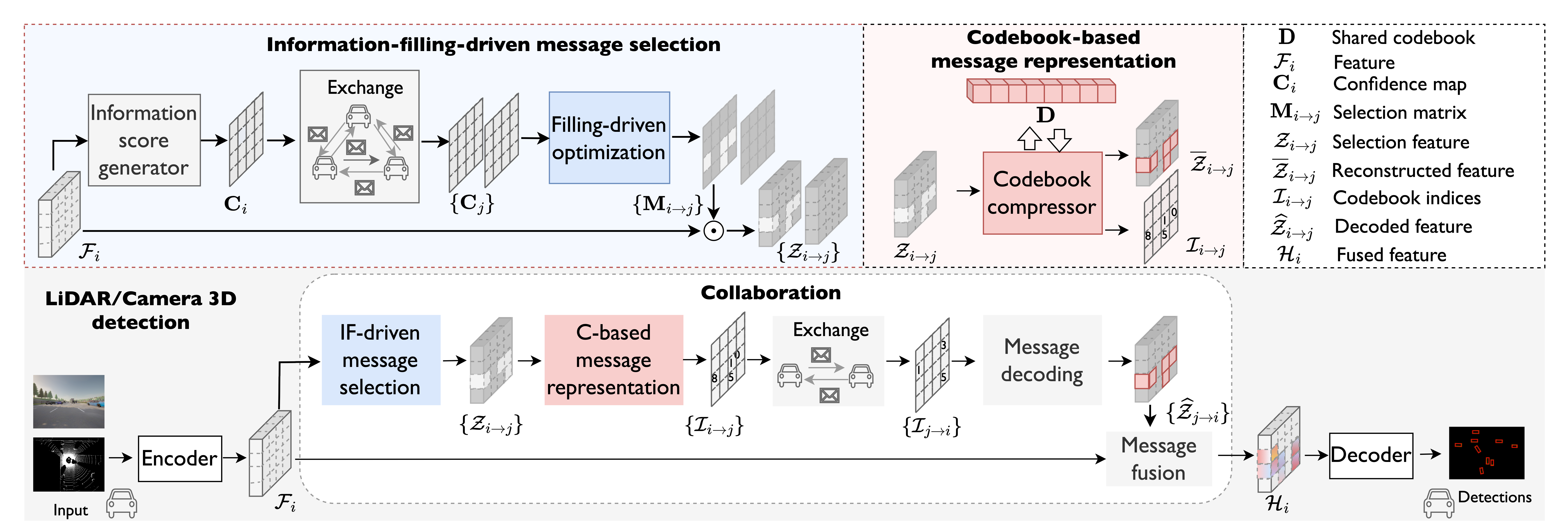}
  \vspace{-8mm}
  \caption{\texttt{CodeFilling} is a novel communication-efficient collaborative 3D detection system. The proposed information-filling-driven message selection and codebook-based message representation contribute to optimizing collaborative messages.}
  \vspace{-6mm}
  \label{fig:framework}
\end{figure*}

Following the above spirit, we propose~\texttt{CodeFilling}, a novel communication-efficient collaborative 3D detection system; see Figure~\ref{fig:framework}. The proposed~\texttt{CodeFilling} includes four key modules: i) a single-agent detector, providing basic detection capabilities; ii) an novel~\emph{information-filling-driven message selection}, which solves local optimizations for choosing pertinent messages to optimally fill other agents' information demands without causing information flow; iii) an novel~\emph{codebook-based message representation}, which leverages a task-driven codebook to achieve pragmatic approximation of feature maps, enabling the transmission of integer code indices; and iv) a message decoding and fusion module, which integrates the messages to achieve enhanced collaborative detections. 

\texttt{CodeFilling} offers two distinct advantages: 
i) it delivers a substantial advancement in the perception-communication trade-off through the transmission of non-redundant code indices; and ii) it is inclusive to both settings of homogeneous and heterogeneous agents by leveraging a standardized code representation; that is, feature maps obtained from various perception models and diverse sensors can be aligned to the unified feature space provided by the proposed codebook.

To evaluate~\texttt{CodeFilling}, we conduct extensive experiments on real-world dataset DAIR-V2X, and a new simulation dataset OPV2VH+ under two homogeneous (LiDAR, camera) and one heterogeneous setting. The results show that i) \texttt{CodeFilling} achieves superior performance performances than Where2comm, the current SOTA, with 1333/1206 $\times$ less communication cost on DAIR-V2X/OPV2VH+; and ii) \texttt{CodeFilling} maintains superior trade-off in both homogeneous and heterogeneous settings, establishing an inclusive collaboration system.

To sum up, our main contributions are three-fold:

$\bullet$ We propose~\texttt{CodeFilling}, a novel communication-efficient collaborative 3D detection system, which significantly improves the perception-
communication trade-off and is inclusive to both homogeneous and heterogeneous collaboration settings;

$\bullet$ We propose two novel methods to optimize collaborative messages: codebook-based message representation and information-filling-driven message selection;

$\bullet$ We conduct comprehensive experiments to validate that \texttt{CodeFilling} achieves SOTA perception-communication trade-off across varying communication bandwidths, on both real-world and simulation datasets in both homogeneous and heterogeneous settings.

\vspace{-2mm}
\section{Related works}
\label{sec:related_work}
\vspace{-2mm}

\textbf{Collaborative Perception.} Collaborative perception~\cite{LiuWho2com:ICRA20,LiuWhen2com:CVPR20,WangV2vnet:ECCV20,XuOPV2V:ICRA22,LiLearning:NeurIPS21,LiV2XSim:RAL22,YuDAIRV2X:CVPR22,XuCoBEVT:CoRL22,HuWhere2comm:NeurIPS22,LuRobust:ICRA23,HuCollaboration:CVPR23,XiangV2XPASG:ICRA22} is an emerging application of multi-agent communication systems to perception tasks, which promote the crucial perception module through communication-enabled complementary perceptual information sharing.
Several high-quality datasets have emerged~\cite{LiV2XSim:RAL22,XuOPV2V:ICRA22,YuDAIRV2X:CVPR22,HuWhere2comm:NeurIPS22,HuCollaboration:CVPR23,YuV2XSeq:CVPR23} to aid in the algorithm development.
Collaborative perception systems have made remarkable progress in improving perception performance~\cite{XuCoBEVT:CoRL22,XuV2XViT:ECCV22} and robustness on practical issues, such as communication bandwidth constraints, pose error~\cite{VadiveluLearning:CoRL20,LuRobust:ICRA23} and latency~\cite{LeiLatency:ECCV22,CoBEVFlow:NeurIPS23}. 
Here, considering that communication efficiency is the bottleneck issue for the scale-up of collaborative perception, we aim to optimize the performance-communication trade-off instead of solely promoting the perception performance regardless of bandwidth costs.

\noindent\textbf{Communication efficiency in collaborative perception.}
To address this bottleneck issue, prior methods have made efforts in two key aspects: message selection and message representation.
For message selection, When2com~\cite{LiuWhen2com:CVPR20} and Who2comm~\cite{LiuWho2com:ICRA20} employ a handshake mechanism, to select information from all the relevant collaborators. 
Furthermore, Where2comm~\cite{HuWhere2comm:NeurIPS22} and CoCa3D~\cite{HuCollaboration:CVPR23} expand the selection process to incorporate spatial dimensions.
For message representation, intermediate feature representation~\cite{WangV2vnet:ECCV20,LiLearning:NeurIPS21,HuWhere2comm:NeurIPS22,LiV2XSim:RAL22,XuV2XViT:ECCV22,XuCoBEVT:CoRL22,HuCollaboration:CVPR23} has demonstrated a more balanced performance-communication trade-off. Source coding~\cite{WangV2vnet:ECCV20} and channel compression~\cite{LiLearning:NeurIPS21} techniques are used to further enhance feature representation efficiency.
However, previous methods accumulate redundant information from various collaborators and still transmit high-dimensional feature vectors, incurring high communication costs. 
Here, we facilitate essential supportive information exchange among agents with compact codebook-based representation, efficiently enhancing detection performance.

\noindent\textbf{Codebook compression.} 
Codebook compression, a lossless compression technique, effectively captures the essence of high-dimensional vectors through the combination of these codes~\cite{GershoVector:1991}. It has diverse applications, ranging from digital image compression~\cite{GershoImage:ICASS1982} to the neural network parameters compression~\cite{HanDeep:CVPR2015}. 
Recently, task-adaptive codebooks have emerged. Rather than pursuing lossless compression, it drops task-irrelevant information and focuses on essential information for specific downstream tasks, further improving representation efficiency~\cite{ChenDianNao:ICASP2014,SinghE2E:ICIP2020}.
However, existing task-adaptive codebooks have largely concentrated on 2D classification tasks~\cite{SinghE2E:ICIP2020}. 
Here, we explore the novel realm of collaborative 3D object detection, introducing fresh challenges for codebook compression. This entails preserving an extensive feature set for precise 3D interpretation and adapting to fluctuating communication bandwidths, necessitating versatile codebook configurations.


\vspace{-2mm}
\section{Problem Formulation}
\label{sec:formulation}
\vspace{-1mm}

Consider $N$ homogeneous or heterogeneous agents in the scene, each has its individual perceptual task and unique sensor setup. To enhance the perception abilities, the agents exchange complementary perceptual information, forming a decentralized mutually beneficial collaboration network. Each agent concurrently acts as both a supporter and a receiver. In their role as supporters, they contribute perceptual information to assist their counterparts. Conversely, as receivers, they gain from the messages provided by others. Such collaborative perception leads to a holistic enhancement of perceptual capabilities. Here we focus on 3D object detection. Let $\mathcal{X}_i$ be the input collected by the $i$th agent's sensor (LiDAR or camera), and $\mathcal{O}^0_i$ be the corresponding ground-truth detection.  The objective is to maximize the detection performances of all agents given certain communication budget B; that is,
\begin{equation}
\small
    \underset{\theta,\mathcal{P}}{\max}~\sum_{i} 
     g \left(\Phi_{\theta} \left(\mathcal{X}_i,\{\mathcal{P}_{j\rightarrow i} \}_{j=1}^N \right), \mathcal{O}^0_i  \right),
    {\rm s.t.~} \sum_{i,j,j\neq i} b(\mathcal{P}_{j\rightarrow i}) \leq B,\label{eq:formulation}
\end{equation}
where $g(\cdot,\cdot)$ is the detection evaluation metric, $\Phi(\cdot)$ is a detection model with trainable parameter $\theta$, $\mathcal{P}_{j\rightarrow i}$ is the message transmitted from the $i$th agent to the $j$th agent, and $b(\cdot)$ measures the communication cost of the collaborative messages. The key challenge is to determine the messages $\mathcal{P}_{j\rightarrow i}$, which should be both informative and compact.

\vspace{-1mm}
\section{CodeFilling: Collaborative 3D Detection}
\label{sec:method}
\vspace{-1mm}
To optimize the trade-off between perception ability and communication cost, we present~\texttt{CodeFilling}, a novel communication-efficient collaborative 3D detection system; see Fig.~\ref{fig:framework}. It has two parts: i) single-agent 3D detection, which allows an agent to equip basic detection ability, implementing $\Phi_{\theta}$ in~\eqref{eq:formulation}, and ii) multi-agent collaboration, enhancing an agent's detection ability through the exchange of efficient perceptual messages $\mathcal{P}$ in~\eqref{eq:formulation}. 

\vspace{-1mm}
\subsection{Single-agent 3D detection}\label{subsec:detector}
\vspace{-1mm}
An agent learns to detect 3D objects based on its sensor inputs. It involves an observation encoder and a detection decoder.~\texttt{CodeFilling} allows agents to accept multi-modality inputs, including RGB images and 3D point clouds. Each agents with its distinct modality projects its perceptual information to the unified global bird's eye view (BEV) coordinate system, better supporting inter-agent collaboration and more compatible with both homogeneous and heterogeneous settings.

\textbf{Observation encoder.} The observation encoder extracts feature maps from the sensor data. For the $i$th agent, given its input $\mathcal{X}_i$, the BEV feature map is 
$ 
\mathcal{F}_i = \Phi_{\rm enc}(\mathcal{X}_i) \in \mathbb{R}^{H \times W \times C},$
where $\Phi_{\rm enc}(\cdot)$ is the encoder and $H,W,C$ are its height, weight and channel. For image inputs, $\Phi_{\rm enc}(\cdot)$ is followed by an additional warping function that transforms the extracted front-view feature to BEV. The BEV feature is output to the decoder, and also message selection and fusion modules when collaboration is established.

\textbf{Detection decoder.} The detection decoder decodes features into objects, including class and regression output. 
Given the feature map $\mathcal{F}_i$, the detection decoder $\Phi_{\rm dec}(\cdot)$ generate the detections of $i$th agent by
$
     \mathcal{O}_i  =  \Phi_{\rm dec}(\mathcal{F}_i) \in \mathbb{R}^{H \times W \times 7},
$
where each location of $\mathcal{O}_i$ represents a rotated box with class $(c,x,y,h,w, \cos\alpha, \sin\alpha)$, denoting class confidence, position, size and angle. 

\vspace{-1mm}
\subsection{Multi-agent collaboration}
\vspace{-1mm}

In the proposed multi-agent collaboration, each agent acts in the dual role of supporter and receiver. As a supporter, each agent employs two novel modules, including information-filling-driven message selection and codebook-based message representation, to determine compact, yet supportive collaboration messages to help others. These two proposed modules enhance communication efficiency in both spatial and channel dimensions of a feature map, respectively. As a receiver, each agent employs a message decoding and fusion module to integrate supportive messages from other agents, improving its perceptual performance.

\vspace{-2mm}
\subsubsection{Information-filling-driven message selection}\label{subsec:selection}
\vspace{-1mm}


To efficiently select compact collaborative messages that support other agents, each agent employs a novel information-filling-driven message selection method. The key idea is to enable each agent to restrainedly select pertinent messages to share with other agents; then collectively, these pieces of non-redundant messages mutually fulfill each other's information demands. For example, in occlusion scenarios, extra information from supporters helps an agent detect missed objects. However, overfilled information from multiple supporters wastes communication resources. Thus, collective coordination is essential to avoid redundancy and enable more beneficial information. To achieve this, the proposed selection has two key steps: information disclosure, wherein agents mutually share their awareness of available information within specific spatial areas, and filling-driven optimization, wherein each agent locally optimizes the supportive messages for others.

\textbf{Information disclosure.}
In information disclosure, each agent: i) employs an information score generator to create its information score map from its feature map, reflecting its available information at each spatial area, and ii) broadcasts this map to all other agents, promoting a mutually thorough awareness of all the available support.

The information score map is implemented with the detection confidence map. Intuitively, the areas containing an object are likely to offer more useful information for revealing missed detections and, therefore, should be assigned higher information scores.
Specifically, given a BEV feature map $\mathcal{F}_i$, its spatial information score map is 
\begin{equation}
\label{eq:generator}
\setlength{\abovedisplayskip}{1pt}
\setlength{\belowdisplayskip}{1pt}
    \mathbf{C}_i  =  \Phi_{\rm generator}(\mathcal{F}_i)  \in [0,1]^{H \times W},
\end{equation}    
where $\Phi_{\rm generator}(\cdot)$ is implemented by detection decoder.  When information score map is generated, each agent broadcasts it to other agents. This initial communication is efficient because of lightweight information score maps.




\textbf{Filling-driven optimization.} 
In the role of a supporter, each agent gathers the other agents' information score maps and determines who needs perceptual information at which spatial areas by locally solving a filling-driven optimization. Here, filling the information demands means that an agent only requires the necessary information at certain spatial areas for precise detection, as extra information no longer provides significant benefits. This requires each supporter to prioritize non-redundant and informative spatial regions with higher scores to assist others and halts the selection once the receiver's information demands are fulfilled.

\begin{figure}[!t]
    \centering
    \includegraphics[width=1.0\linewidth]{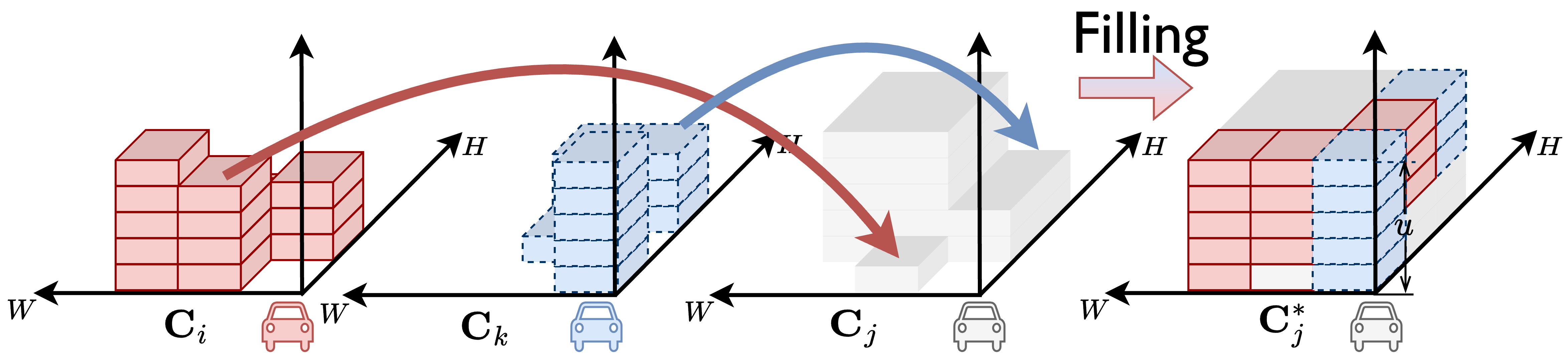}
  \vspace{-7mm}
  \caption{The information-filling-driven message selection fulfills the information demand with non-redundant information.}
  \vspace{-6mm}
  \label{fig:filling}
\end{figure}

Specifically, the optimization is formulated as a proxy-constrained problem and obtains a binary selection matrix for each agent to support each receiver. Let $\mathbf{M}_{i\rightarrow j} \in \{0,1\}^{H \times W}$ be the binary selection matrix supported on the BEV map. Each element in the matrix indicates whether Agent $i$ should send the information to Agent $j$ at a specific spatial location (1 for sending information, and 0 for not sending).  To solve for the binary selection matrix, the proxy-constrained problem is formulated as follows,
\vspace{-2mm}
\begin{subequations}
\small
\setlength{\belowdisplayskip}{1pt}
\setlength{\jot}{-12pt}
    \begin{align}
        &\{ \mathbf{M}_{i \rightarrow j}^{*} \}_{i,j} \ = \ \underset{\mathbf{M}}{\rm argmax}\sum_{j=1}^{N} f_{\rm min}\left(\mathbf{C}_j+\sum_{i=1,i\neq j}^{N}\mathbf{M}_{i\rightarrow j}\odot \mathbf{C}_{i},u\right),\label{eq:selector}\\
        &{\rm where}\sum_{i,j=1,j\neq i}^{N}\mathbf{M}_{i\rightarrow j} \leq b,\mathbf{M}_{i\rightarrow j}\in\{0,1\}^{H\times W}\label{eq:constraint}.
    \end{align}
\end{subequations}    
Here $\odot$ denotes element-wise multiplication, and the scalar $u$ is a hyper-parameter to reflect the upper bound of information demand. 
The function $f_{\rm min}(\cdot,\cdot)$ computes the element-wise minimum between a matrix and a scalar.

In~\eqref{eq:selector}, $\mathbf{C}_j+\sum_{i=1,i\neq j}^N\mathbf{M}_{i\rightarrow j}\odot \mathbf{C}_i$ indicates that each receiver $j$ accumulates the information transmitted from all supporters, combined with its own information, at each location, $f_{\rm min}(\mathbf{C}_j+\sum_{i=1,i\neq j}^N\mathbf{M}_{i\rightarrow j}\odot \mathbf{C}_i,u)$ denotes the utility for each receiver, linearly increasing with the accumulated information scores until reaching the information demand $u$. 
Note that: i)~\eqref{eq:selector} is solved at the supporter side for preparing messages to a receiver; ii) the sum-based utility motivates supporters to collectively meet the receiver's demand and focus on higher-scoring regions, and iii) the cutoff point leads to halting selection to prevent redundancy.

Equation~\eqref{eq:selector} transforms the feature-based collaboration utility in~\eqref{eq:formulation} as the sum of the information scores. This is based on the assumption that the accumulation of information scores mirrors the benefits of feature aggregation.
Equation~\eqref{eq:constraint} addresses the bandwidth limitation in~\eqref{eq:formulation} by quantifying the total number of selected regions.
This approach simplifies the objective in~\eqref{eq:formulation} into a proxy-constrained problem in~\eqref{eq:selector} and ~\eqref{eq:constraint}. The optimized selection solution derived from~\eqref{eq:formulation} is expected to yield a superior outcome in the final feature-based collaboration.




This optimization problem has an analytical solution; see the theoretical derivation in the appendix. The solving process incurs a computational cost of $O(\log(m))$, where $m$ denotes the number of spatial region candidates. By focusing on the extremely sparse foreground areas, we effectively reduce the cost to a negligible level, enabling each agent to provide more targeted support for others with minimal cost.


Based on the optimized selection matrix $\{\mathbf{M}_{i\rightarrow j}^{*}\}_{j=1}^{N}$, each agent supports each collaborator with a sparse yet informative feature map $\mathcal{Z}_{i\rightarrow j}=\mathbf{M}_{i\rightarrow j}^{*}\odot \mathcal{F}_i$, promising superior perception improvements given the limited communication budget. These selected sparse feature maps are then output to the message representation module.

The proposed message selection offers two key benefits: i) it avoids redundancy from multiple supporters via collective selection, and ii) it adapts to varying communication conditions by adjusting information demand, lower demand for efficiency in limited budgets, and higher demand for superior performance in ample budgets. 
Compared to existing selection methods~\cite{LiuWhen2com:CVPR20,LiuWho2com:ICRA20,HuWhere2comm:NeurIPS22,HuCollaboration:CVPR23},
which are based on individual supporter-receiver pairs, our collective optimization further reduces redundancy across various supporters.


\vspace{-2mm}
\subsubsection{Codebook-based message representation}\label{subsec:encoding}
\vspace{-1mm}

To efficiently transmit the selected feature map $\mathcal{Z}_{i\rightarrow j}$, each agent leverages a novel codebook-based message representation, reducing communication cost along the channel dimension. The core idea is to approximate a high-dimensional feature vector by the most relevant code from a task-driven codebook; as a result, only integer code indices need to be transmitted, rather than the complete feature vectors composed of floating-point numbers.


\textbf{Codebook learning.} Analogous to a language dictionary used by humans, our task-driven codebook is shared among all agents to standardize their communication for achieving the detection task. This codebook consists of a set of codes, which are learned to pragmatically approximate possible perceptual features present in the training dataset. Here the pragmatic approximation refers to each code serving as a lossy approximation of a feature vector, while retaining essential information necessary for the downstream detection task within that vector. Specifically, let $\digamma=\{\mathcal{F}^{(i,s)}\}_{i=1,s=1}^{N,S}$ be the collective set of BEV feature maps extracted by the observation encoders of all $N$ agents across all $S$ training scenes. Let $\mathbf{D} =   \begin{bmatrix} \mathbf{d}_1, \mathbf{d}_2, \cdots, \mathbf{d}_{n_L} \end{bmatrix}
\in\mathbb{R}^{C \times n_L}$ be the codebook, where $\mathbf{D}_{[\ell]} = \mathbf{d}_{\ell} \in \mathbb{R}^C$ is the $\ell$th code and $n_L$ is the number of codes.

The task-driven codebook is learned through feature approximation at each spatial location; that is,
\begin{small}
\begin{eqnarray}
\label{eq:codebook_learning}
\setlength{\abovedisplayskip}{1pt}
\setlength{\belowdisplayskip}{1pt}
\mathbf{D}^*= \arg\min_{\mathbf{D}} \sum_{\mathcal{F}\in\digamma}
\sum_{h,w} \min_{{\ell}} \left(  \Psi( \mathbf{D}_{[\ell]} )  + \left\| \mathcal{F}_{[h,w]} - \mathbf{D}_{[\ell]} \right\|_2^2 \right),
\end{eqnarray}
\end{small}where $\Psi(\cdot)$ denotes the resulting detection performance achieved by substituting $\mathbf{D}_{[\ell]}$ for $\mathcal{F}_{[h,w]}$. The first term pertains to the requirements of the downstream detection task and the second term reflects the reconstruction error between the original feature vector and the code. This approximation is lossy for reconstruction while lossless for the perceptual task, enabling the reduction of communication cost without sacrificing perceptual capacity.




\begin{figure}[!t]
    \centering
    \includegraphics[width=1.0\linewidth]{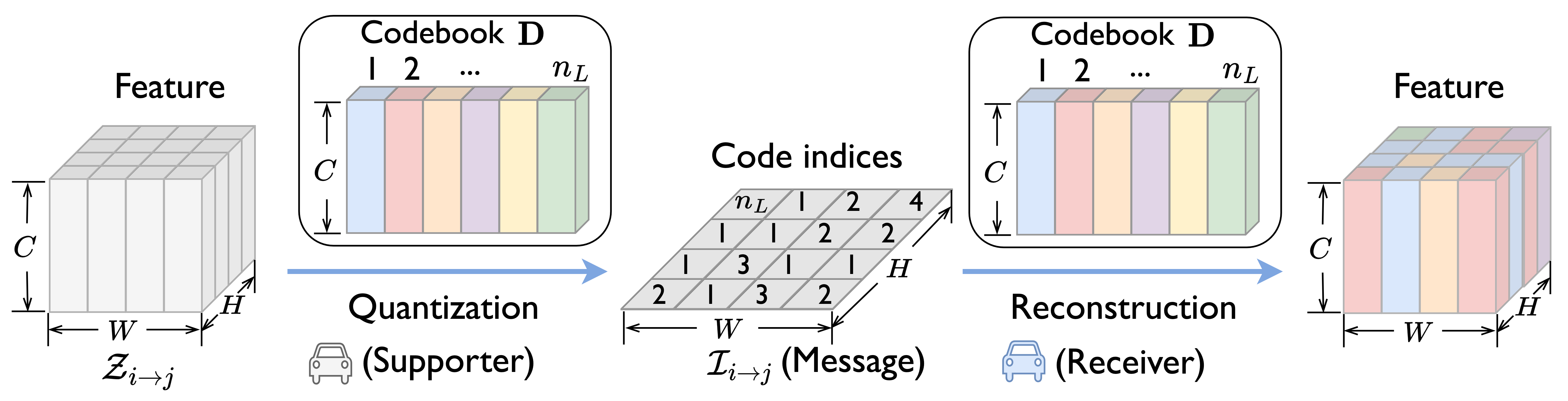}
  \vspace{-8mm}
  \caption{The codebook-based message representation depicts the original feature vector with the most relevant codes.}
  \vspace{-6mm}
  \label{fig:codebook}
\end{figure}

\textbf{Code index representation.} 
Based on the shared codebook $\mathbf{D}$, each agent can substitute the selected sparse feature map $\mathcal{Z}_{i\rightarrow j}$ by a series of code indices $\mathcal{I}_{i\rightarrow j}$. For each BEV location $(h,w)$, the code index is obtained as,
\vspace{-1mm}
\begin{equation}
\label{eq:code_index}
\setlength{\abovedisplayskip}{1pt}
\setlength{\belowdisplayskip}{1pt}
\small
{(\mathcal{I}_{i\rightarrow j})}_{[h,w]} = \arg \min_{\ell} \left\| {(\mathcal{Z}_{i\rightarrow j})}_{[h,w]} - \mathbf{D}_{[\ell]}\right\|_2^2.
\end{equation} 
The codebook offers versatility in its configuration by adjusting both the codebook size $n_L$ and the quantity of codes $n_R$ used for representing the input vector. Equation~\eqref{eq:code_index} demonstrates a specific instance where $n_R=1$, chosen for simplicity in notation. When $n_R$ is larger, the representation involves a combination of multiple codes.

Overall, the final message sent from the $i$th agent to the $j$th agent is $\mathcal{P}_{i\rightarrow j}=\mathcal{I}_{i\rightarrow j}$, conveying the required complementary information with compact code indices. Agents exchange these packed messages with each other.


This codebook-based representation offers three advantages: i) efficiency for transmitting lightweight code indices; ii) adaptability to various communication resources via adjusting code configurations (smaller for efficiency, larger for superior performance), and iii) extensibility by providing a shared standardized representation. New heterogeneous agents can easily join the collaboration by adding its effective perceptual feature basis to the codebook.



\begin{figure*}[!t]
  \centering
  \begin{subfigure}{0.3\linewidth}
    \includegraphics[width=0.9\linewidth]{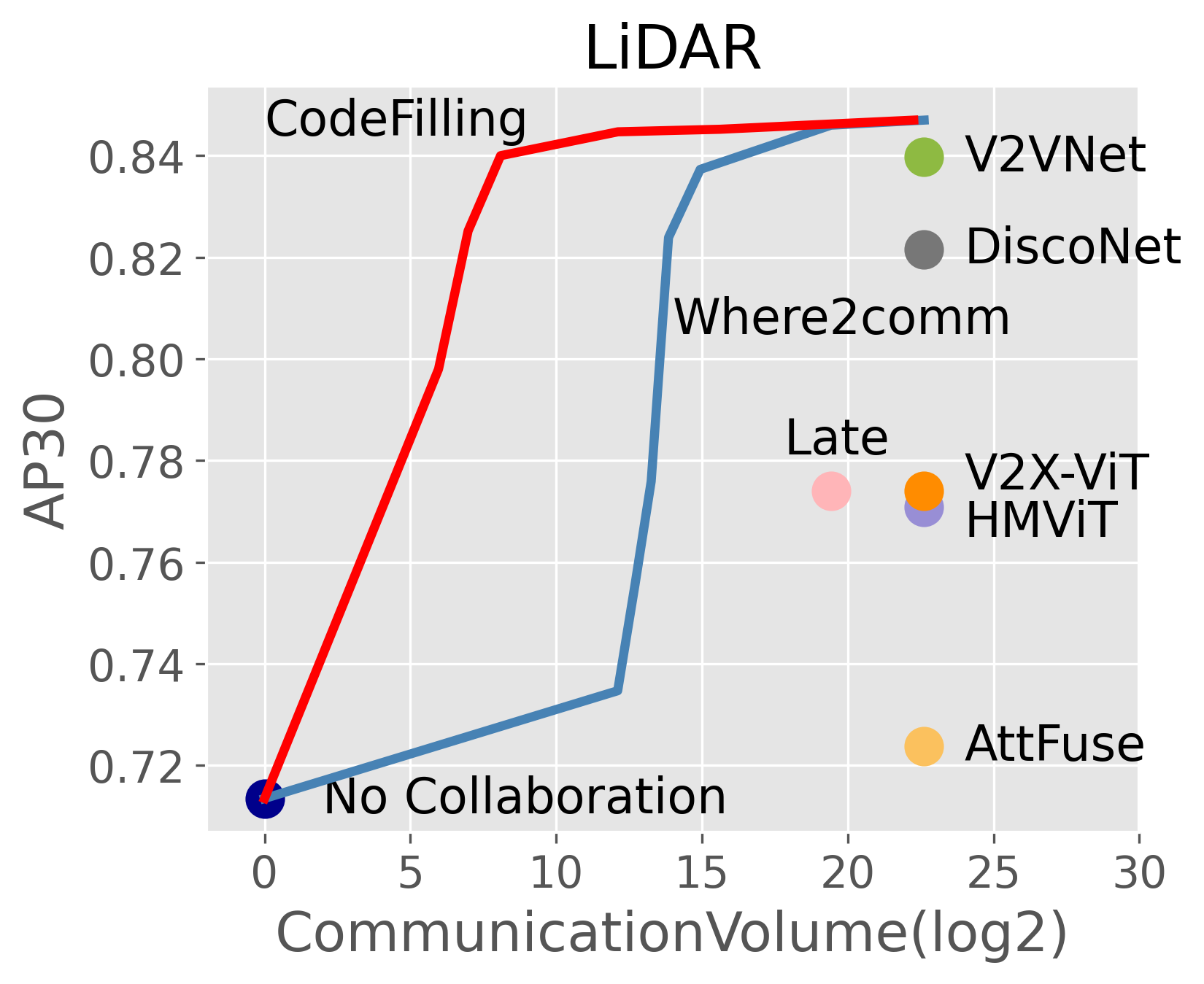}
  \end{subfigure}
  \begin{subfigure}{0.3\linewidth}
    \includegraphics[width=0.9\linewidth]{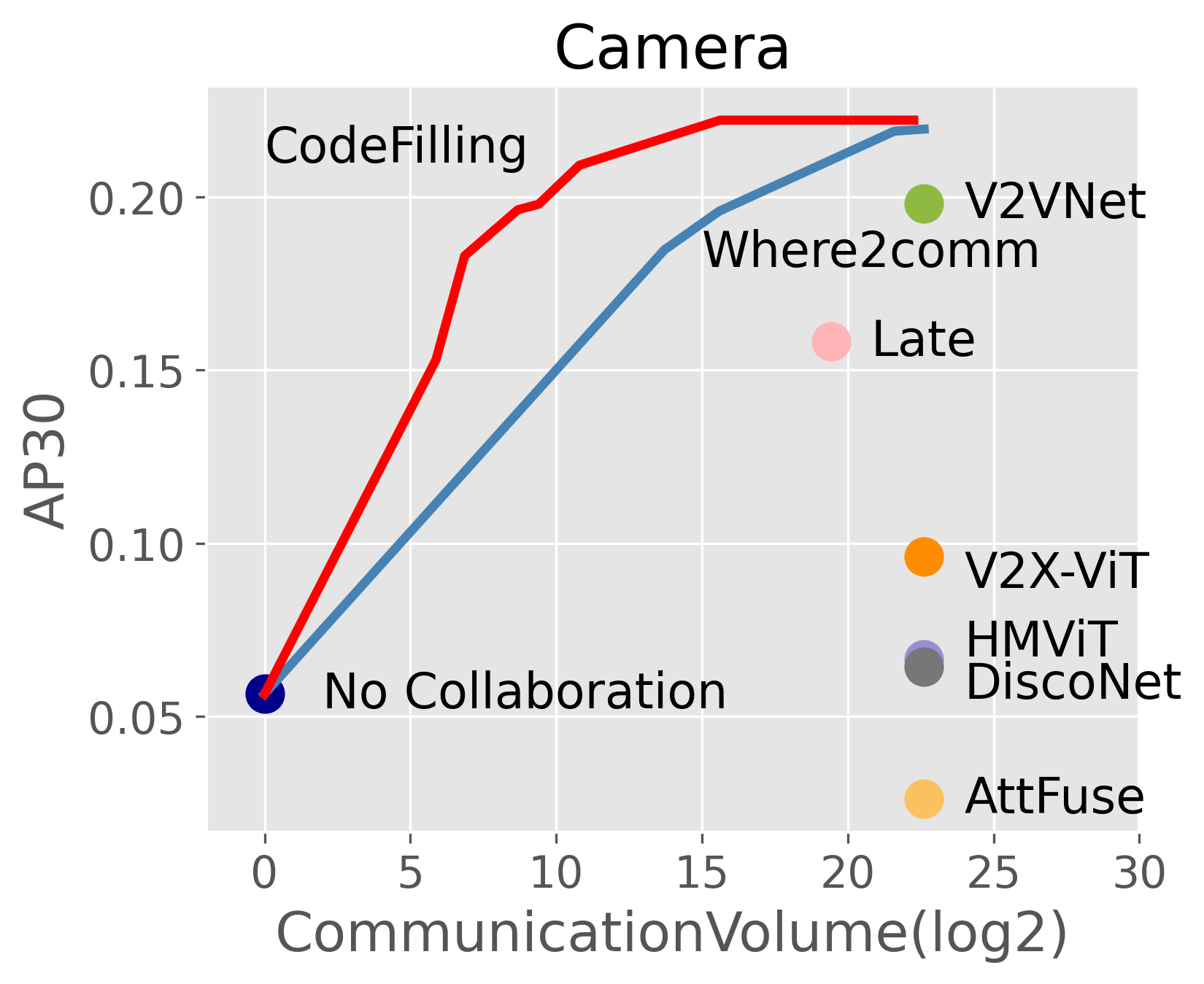}
  \end{subfigure}
  \begin{subfigure}{0.3\linewidth}
    \includegraphics[width=0.9\linewidth]{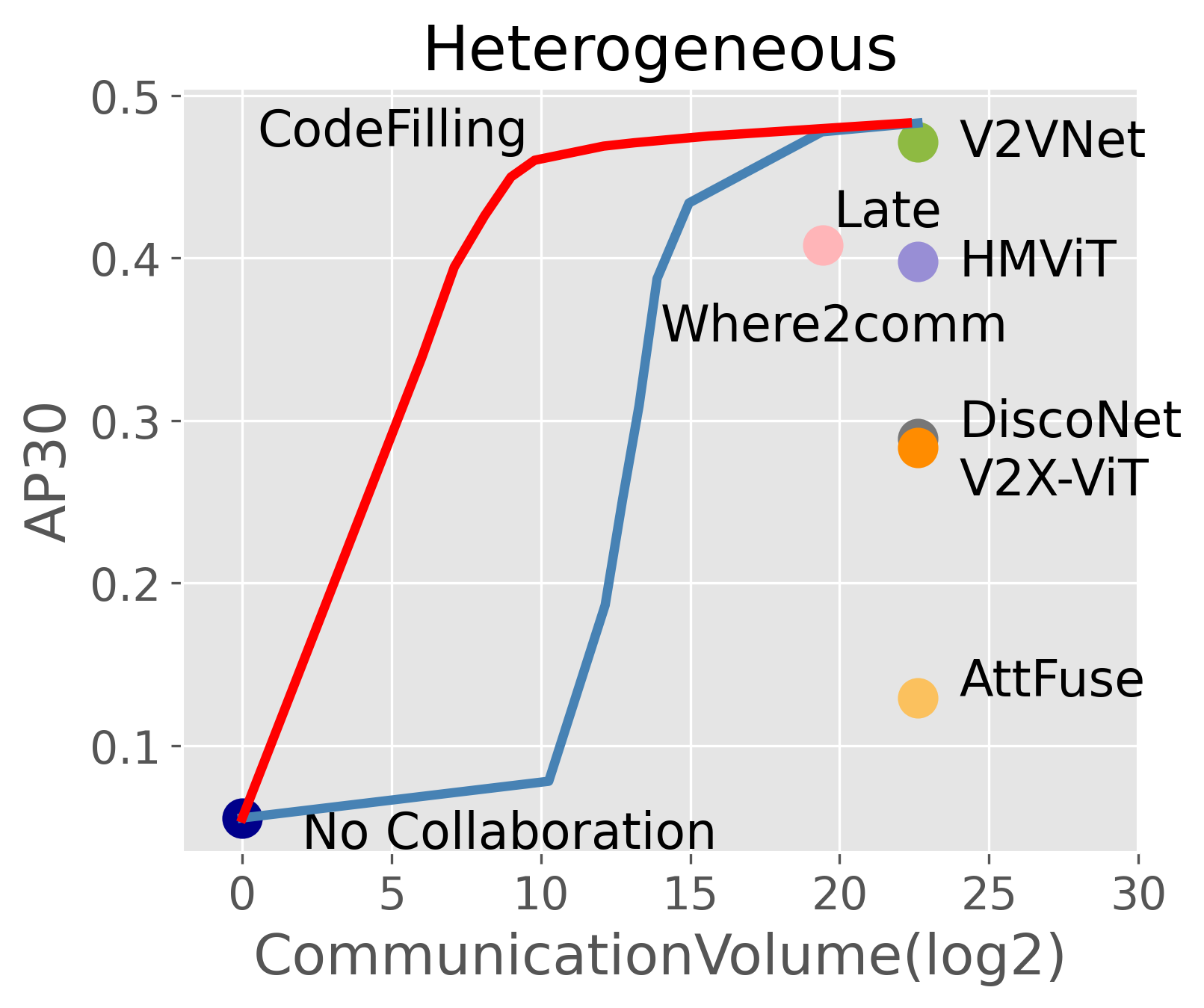}
  \end{subfigure}
  \vspace{-4mm}
  \caption{In DAIR-V2X, \texttt{CodeFilling} achieves the best perception-communication trade-off in~\emph{homogeneous \& heterogeneous} settings.}
  \vspace{-5mm}
  \label{Fig:DAIRV2X_SOTA}
\end{figure*}
\begin{figure*}[!t]
  \centering
  \begin{subfigure}{0.3\linewidth}
    \includegraphics[width=0.9\linewidth]{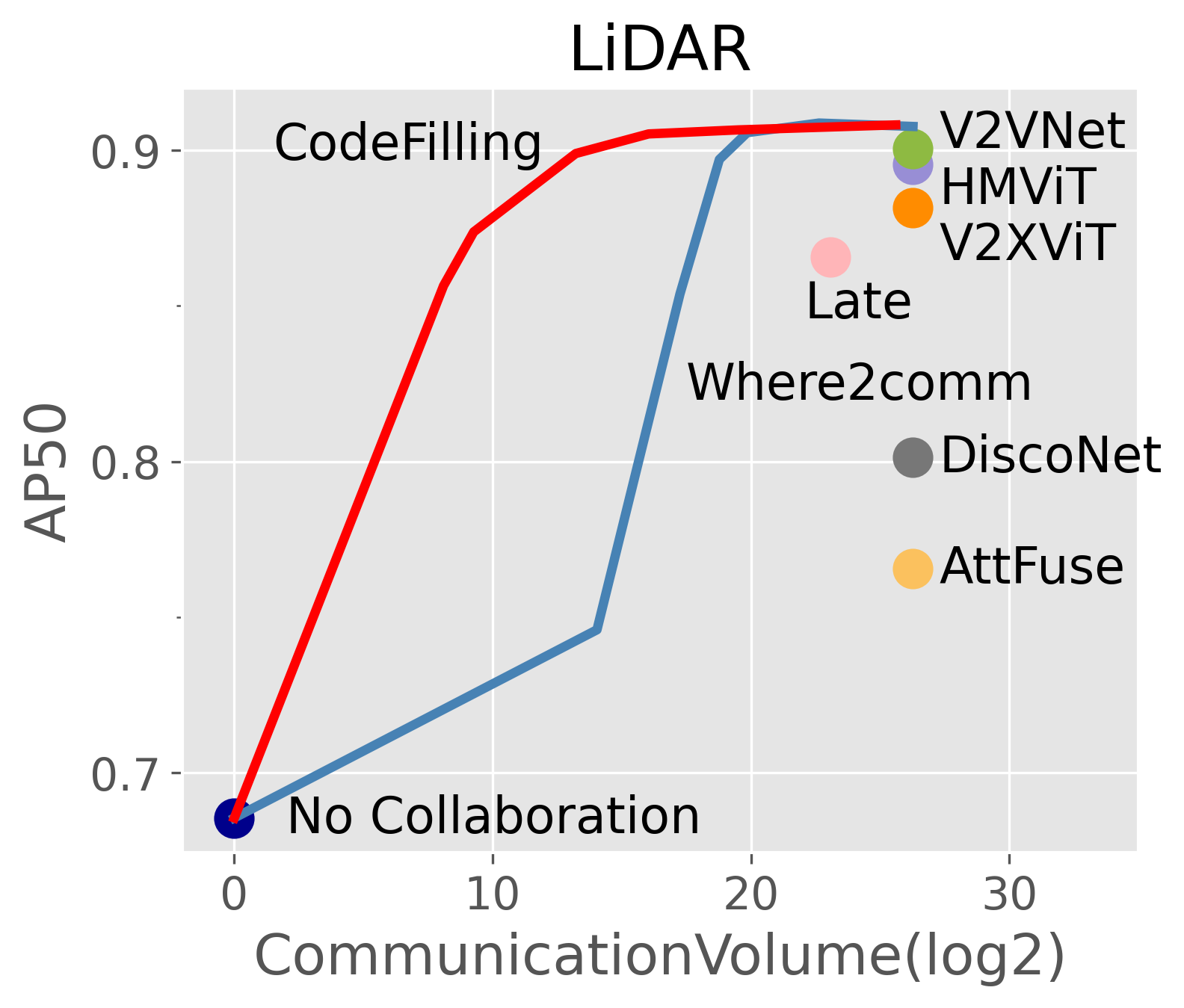}
  \end{subfigure}
  \begin{subfigure}{0.3\linewidth}
    \includegraphics[width=0.9\linewidth]{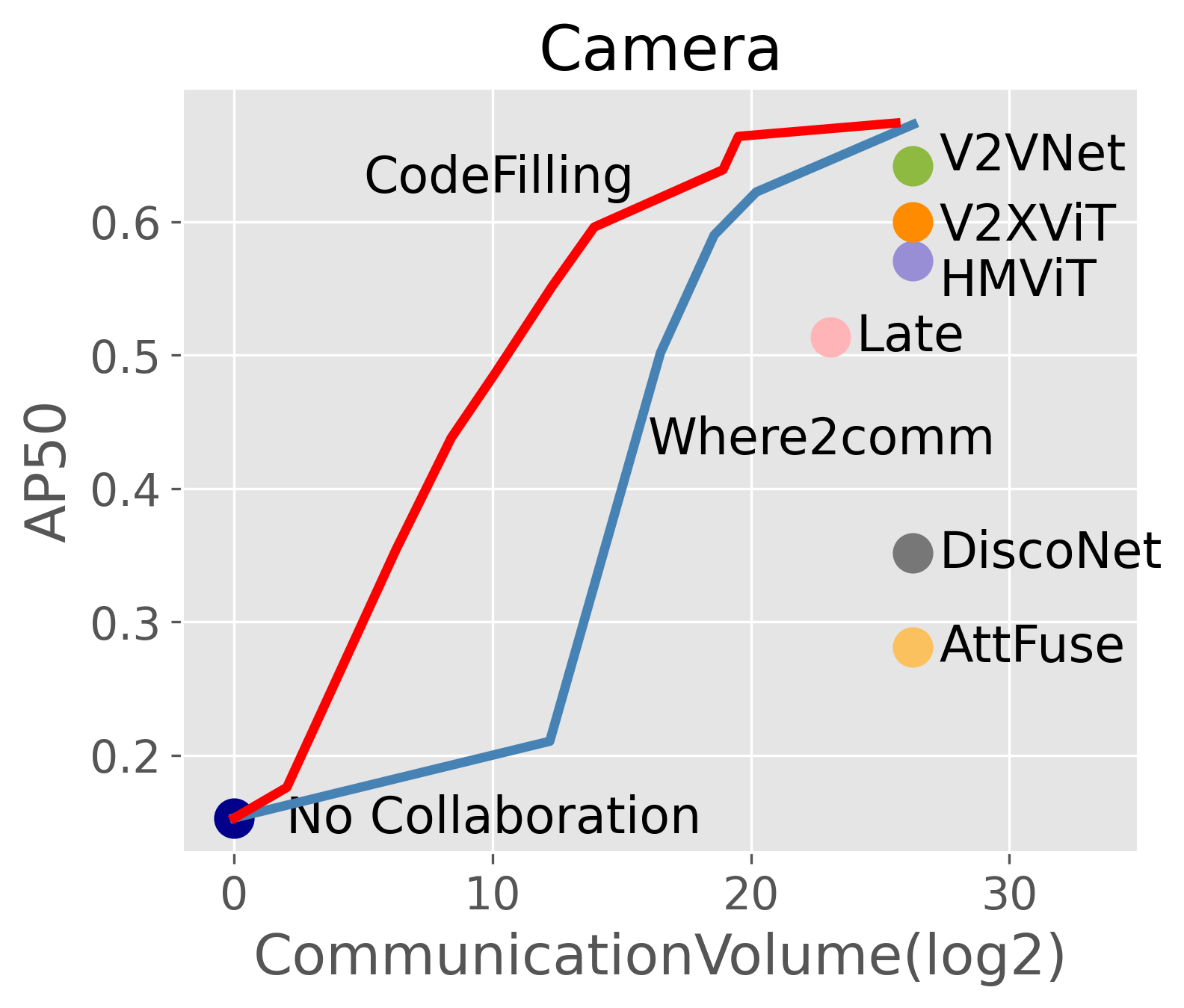}
  \end{subfigure}
  \begin{subfigure}{0.30\linewidth}
    \includegraphics[width=0.9\linewidth]{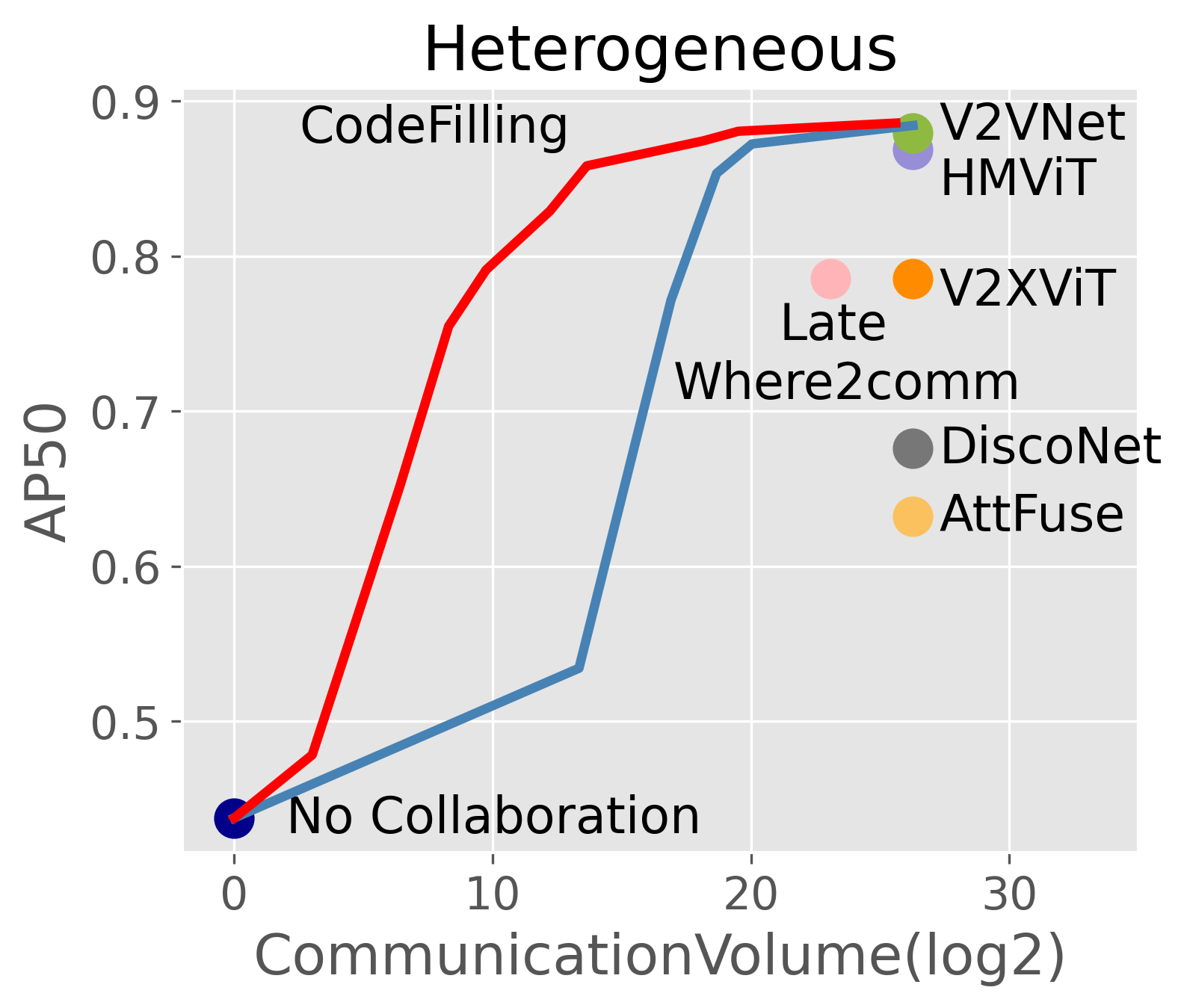}
  \end{subfigure}
  \vspace{-4mm}
  \caption{In OPV2VH+, \texttt{CodeFilling} achieves the best perception-communication trade-off in~\emph{homogeneous \& heterogeneous} settings.}
  \vspace{-5mm}
  \label{Fig:OPV2V_SOTA}
\end{figure*}

 \vspace{-2mm}
\subsubsection{Message decoding and fusion }\label{subsec:decoding}
\vspace{-1mm}
Message decoding reconstructs the supportive features based on the received code indices and the shared codebook.
Given the received message $\mathcal{P}_{j\rightarrow i}={\mathcal{I}}_{j\rightarrow i}$, the decoded feature map's $\widehat{\mathcal{Z}}_{j\rightarrow i}\in\mathbb{R}^{H\times W\times C}$ element located at $(h,w)$ is $(\widehat{\mathcal{Z}}_{j\rightarrow i})_{[h,w]}=\mathbf{D}_{[\mathcal{I}_{j\rightarrow i}[h,w]]}$. Subsequently, message fusion aggregates these decoded feature maps to augment individual features, implementing by the non-parametric point-wise maximum fusion. For the $i$th agent, given the reconstructed feature $\widehat{\mathcal{Z}}_{j\rightarrow i}$. The enhanced BEV feature is obtained as
$\mathcal{H}_i=\underset{j\in\mathcal{N}_i}{\rm max}(\mathcal{F}_{i},{\widehat{\mathcal{Z}}}_{j\rightarrow i})\in\mathbb{R}^{H\times W\times C}$
where $\mathcal{N}_i$ is $i$-th agent's connected collaborators and ${\rm max}(\cdot)$ maximizes the corresponding features from multiple agents at each individual spatial location. The enhanced feature $\mathcal{H}_i$ is decoded to generate the upgraded detection $\widehat{\mathcal{O}}_{i}$. 

\vspace{-1mm}
\subsection{Loss functions}\label{sec4:loss}
\vspace{-1mm}
To train the overall system, we supervise three tasks: information score map generation, object detection, and codebook learning. The information score map generator reuses the parameters of the detection decoder. The overall loss is defined as, 
$
\label{eq:loss}
L = \sum_i^{N} L_{\text{det}} \left(\widehat{\mathcal{O}}_i,\mathcal{O}^{0}_i \right)+\left\| {\mathcal{F}}_{i} - {\overline{\mathcal{F}}}_{i} \right\|_2^2,$
where $L_{\text{det}}(\cdot)$ denotes the detection loss~\cite{ZhouObjects:Arxiv2019}, $\mathcal{O}^{0}_i$ and $\widehat{\mathcal{O}}_i$ represents the ground-truth and predicted objects,  and ${\mathcal{F}}_{i}$ and ${\overline{\mathcal{F}}}_{i}$ denote the $i$-th agent's original feature map and the one approximated by codes. During the optimization, the network parameters and the codebook are updated simultaneously.

\begin{figure*}[!t]
\vspace{-1mm}
\begin{minipage}{.49\linewidth}
  \centering
  \begin{subfigure}{0.98\linewidth}
    \includegraphics[width=1.0\linewidth]{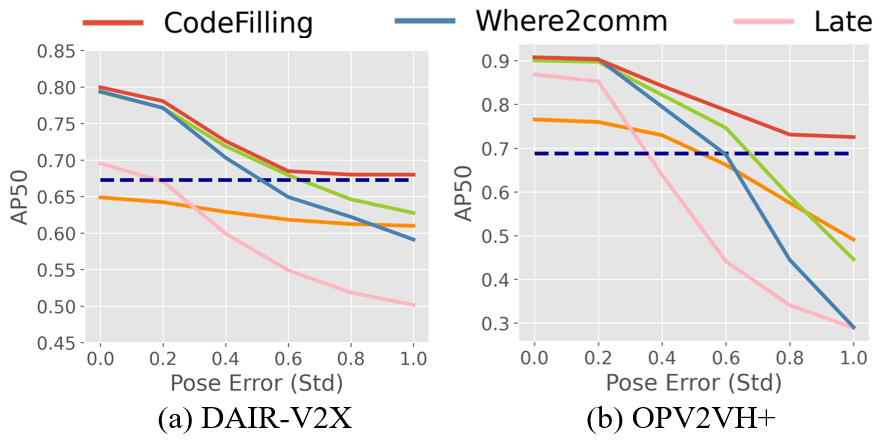}
  \end{subfigure}
  \vspace{-5mm}
  \caption{\texttt{CodeFilling} is robust to pose error issue.}
  \vspace{-3mm}
  \label{Fig:Robust_poseerror}
\end{minipage}
\begin{minipage}{.49\linewidth}
  \centering
   \begin{subfigure}{0.98\linewidth}
    \includegraphics[width=1.0\linewidth]{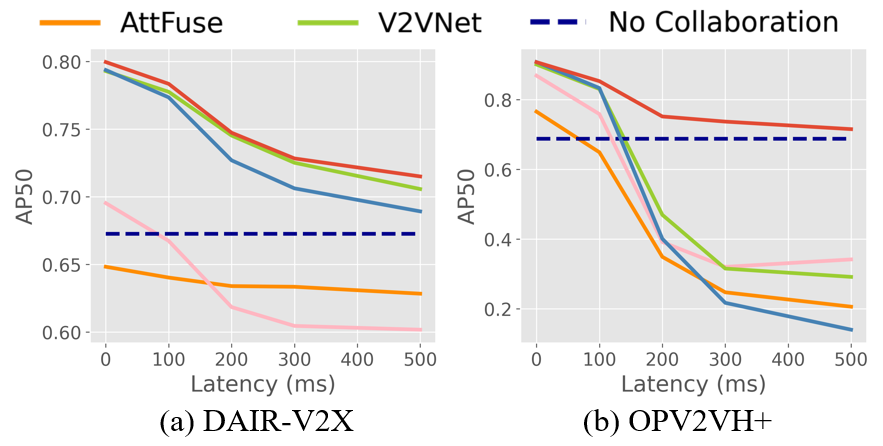}
  \end{subfigure}
  \vspace{-5mm}
  \caption{\texttt{CodeFilling} is robust to communication latency issue.} 
  \vspace{-3mm}
  \label{Fig:Robust_latency}
  \end{minipage}
\end{figure*}

\begin{figure*}[!t]
  \centering
  \begin{subfigure}{0.19\linewidth}
    \includegraphics[width=1.0\linewidth]{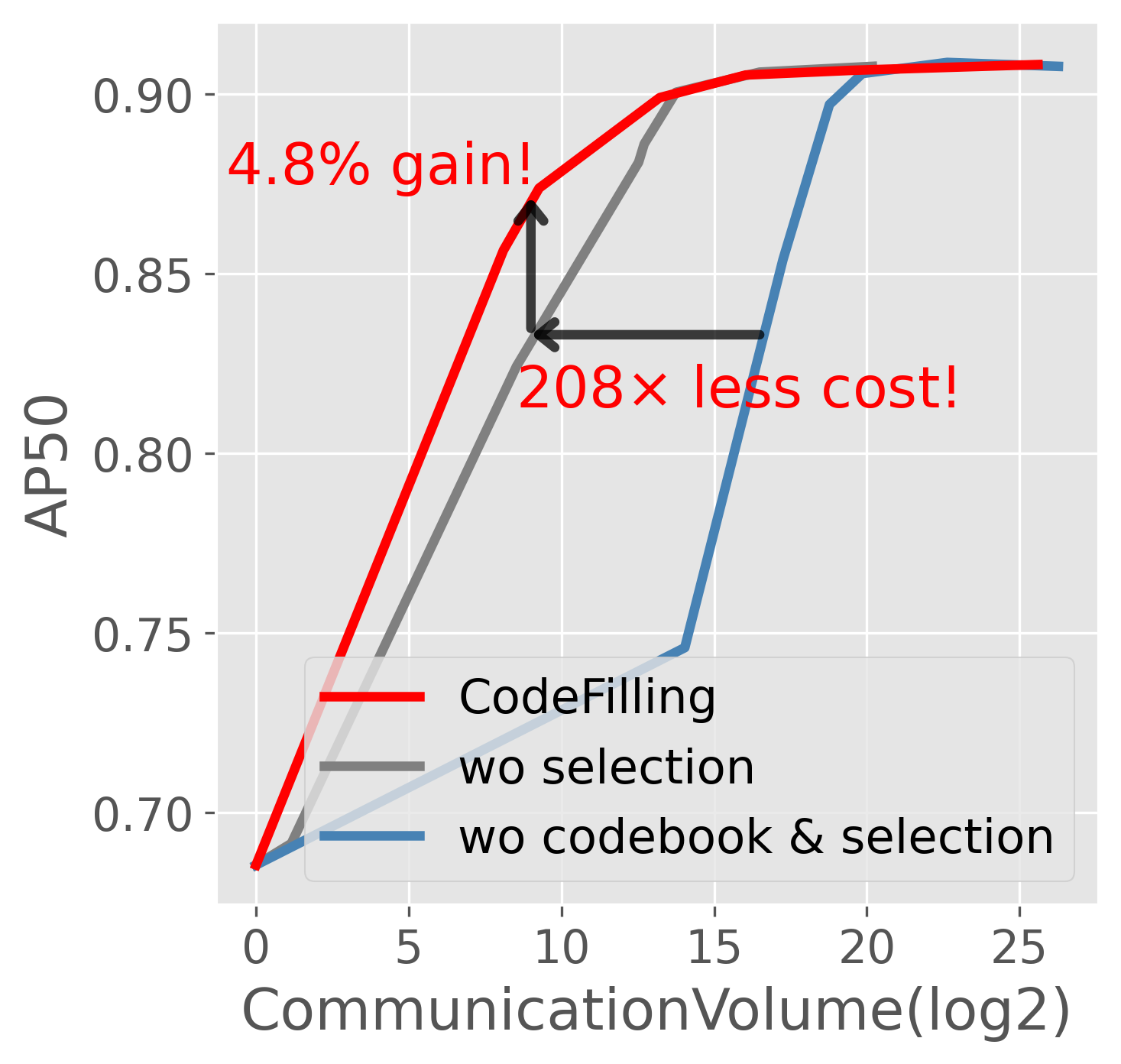}
    \vspace{-5mm}
    \caption{Two components.}
    \label{Fig:Abl_comp}
  \end{subfigure}
  \begin{subfigure}{0.19\linewidth}
    \includegraphics[width=1.0\linewidth]{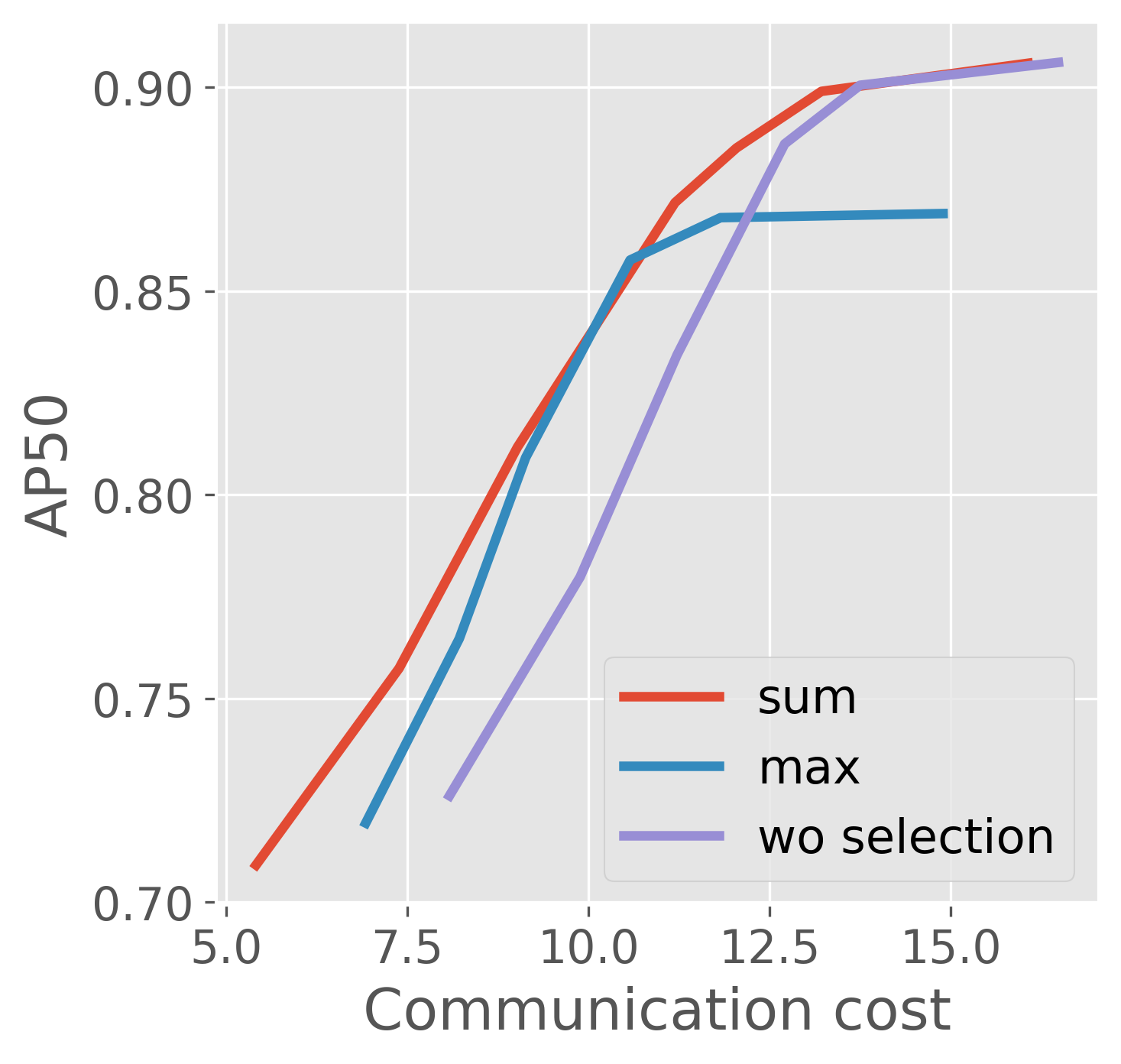}
    \vspace{-5mm}
    \caption{Selection utility design.}
    \label{Fig:Abl_util}
  \end{subfigure}
  \begin{subfigure}{0.19\linewidth}
    \includegraphics[width=1.0\linewidth]{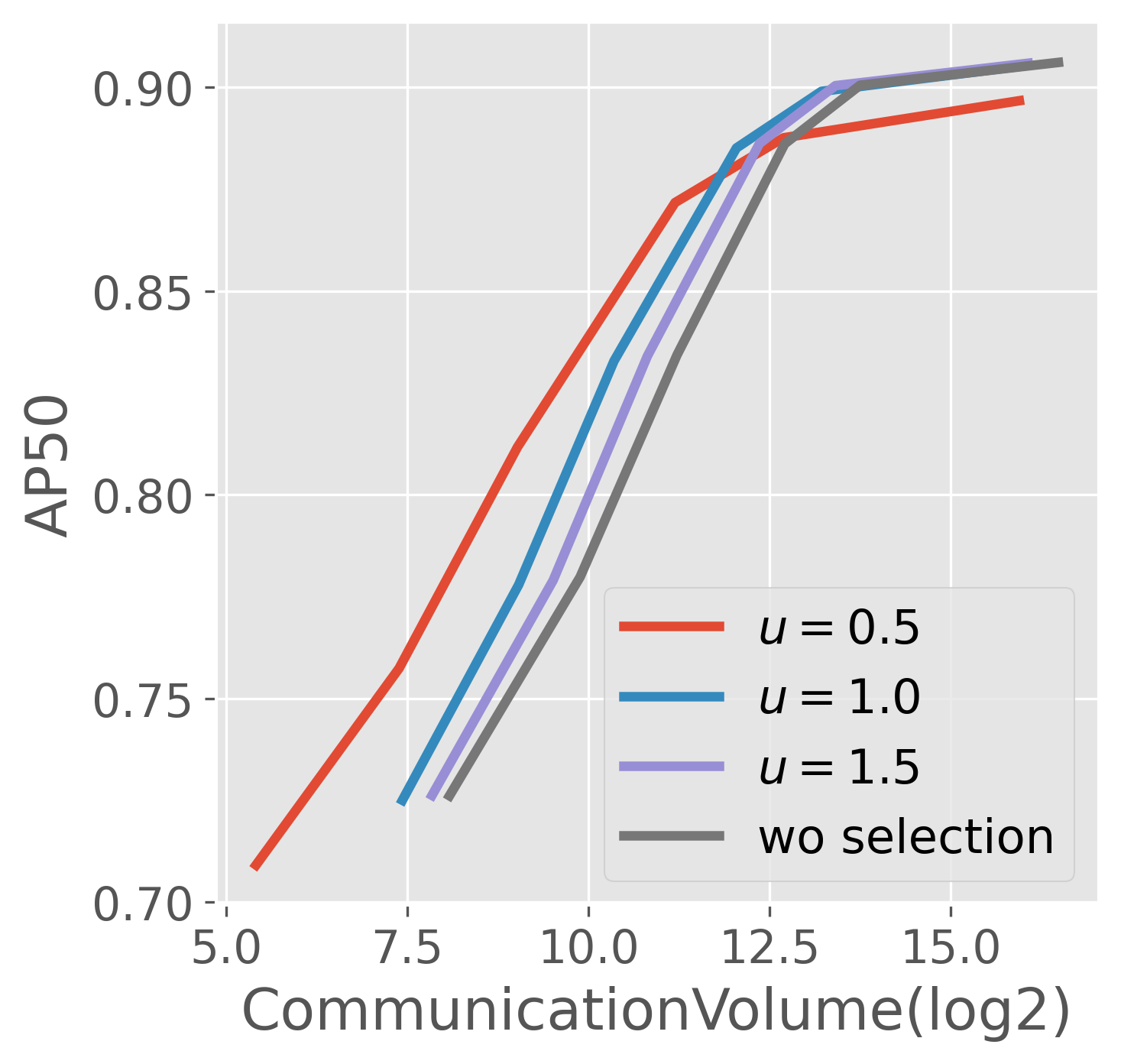}
    \vspace{-5mm}
    \caption{Information demand.}
    \label{Fig:Abl_info}
  \end{subfigure}
  \begin{subfigure}{0.19\linewidth}
    \includegraphics[width=1.0\linewidth]{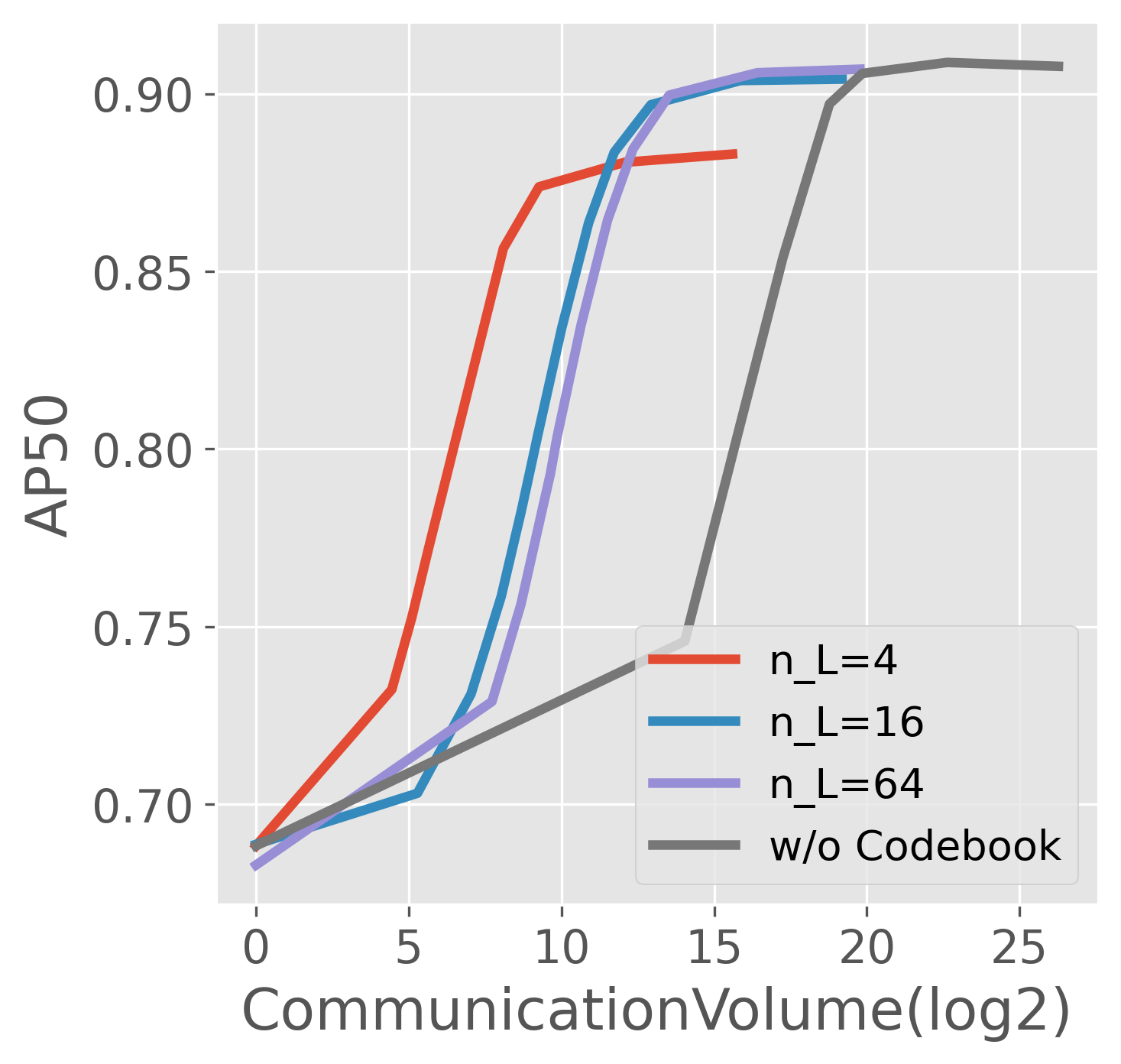}
    \vspace{-5mm}
    \caption{Codebook size.}
    \label{Fig:Abl_size}
  \end{subfigure}
  \begin{subfigure}{0.19\linewidth}
    \includegraphics[width=1.0\linewidth]{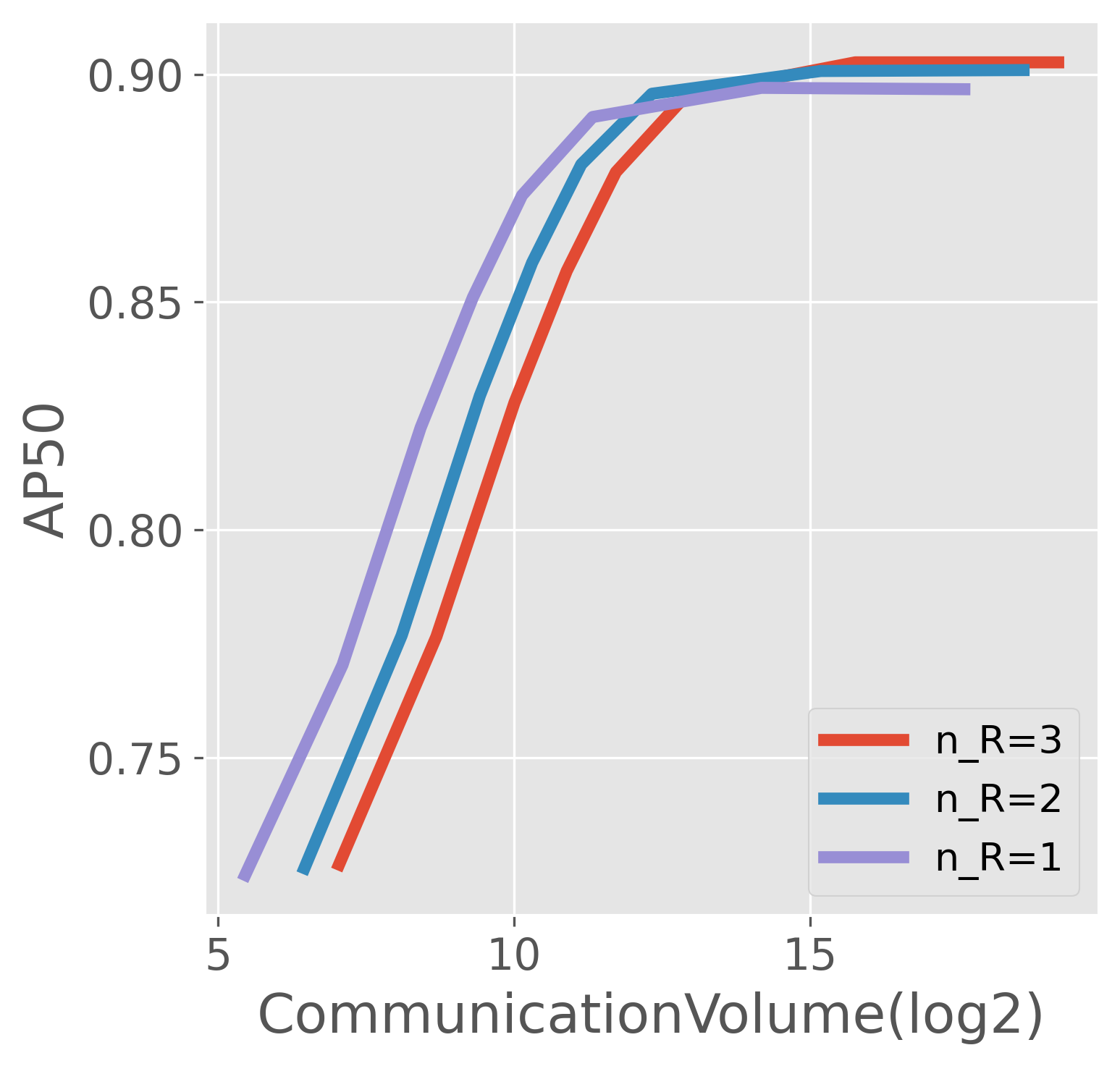}
    \vspace{-5mm}
    \caption{Code quantity.}
    \label{Fig:Abl_residual}
  \end{subfigure}
  \vspace{-3mm}
  \caption{Both the proposed information-filling-driven message selection and codebook-based representation are effective.}
  \vspace{-7mm}
  \label{Fig:Abl_OPV2V}
\end{figure*}

 
\vspace{-2mm}
\section{Experimental Results}
\label{sec:experiments}
\vspace{-1mm}
Our experiments cover two datasets, both real-world and simulation scenarios, two types of sensors (LiDAR and cameras), and both homogeneous and heterogeneous settings. 
Specifically, we conduct 3D object detection in the setting of V2X-communication-aided autonomous driving on DAIR-V2X dataset~\cite{YuDAIRV2X:CVPR22} and the extended large-scale OPV2VH+ dataset. The detection results are evaluated by Average Precision (AP) at Intersection-over-Union (IoU) thresholds of $0.30$ and $0.50$. The communication volume follows the standard setting as ~\cite{XuOPV2V:ICRA22,XiangHMViT:ICCV23,HuWhere2comm:NeurIPS22,HuCollaboration:CVPR23} that counts the message size by byte in log scale with base $2$.

\vspace{-2mm}
\subsection{Datasets and experimental settings}\label{subsec:Datasets}
\vspace{-1mm}

\textbf{DAIR-V2X}~\cite{YuDAIRV2X:CVPR22} is a widely-used \textbf{real-world} collaborative perception dataset. Each scene contains two agents: a vehicle and a road-side-unit. Each agent is equipped with a LiDAR and a camera. The perception range is 204.8m$\times$102.4m. 
\textbf{OPV2VH+} is an extended large-scale version of the original vehicle-to-vehicle camera-only collaborative perception dataset OPV2V+~\cite{HuCollaboration:CVPR23} with a larger array of collaborative agents (a total of $10$) and additional LiDAR sensors, co-simulated by OpenCDA~\cite{XuOpenCDA:ITSC2021} and CARLA~\cite{DosovitskiyCARLA:CoRL2017}. Each agent has a LiDAR, 4 cameras, and 4 depth sensors. The detection range is 281.6m $\times$ 80m.

\noindent
\textbf{Implementation.} We adopt PointPillar~\cite{LangPointPillars:CVPR2018} and CaDDN~\cite{ReadingCategorical:CVPR2021} for the LiDAR and camera detector, respectively. Regarding the heterogeneous setup, agents are randomly assigned either LiDAR or camera, resulting in a balanced 1:1 ratio of agents across the different modalities.

\noindent\textbf{Communication volume.} Specifically, for feature representation, given a selection matrix $\mathbf{M}$, the bandwidth is calculated as ${\rm log}_2(H\times W \times |\mathbf{M}|\times C\times 32/8)$. Here, $32$ represents the float32 data type and $8$ converts bits to bytes. For code index representation, given codebook $\mathbf{D}\in\mathbb{R}^{C\times n_L}$, comprised of $n_L$ codes and each vector constructed using $n_R$ codes, the bandwidth given the selection matrix $\mathbf{M}$ is calculated as ${\rm log}_2(H\times W \times |\mathbf{M}|\times {\rm log}_2(n_L)\times n_R /8)$. Here, ${\rm log}_2(n_L)$ signifies the data amount required to represent each code index integer, decided by the codebook size.

\vspace{-2mm}
\subsection{Quantitative evaluation}\label{subsec:Quantitative}
\vspace{-2mm}


\textbf{Benchmark comparison.} Fig.~\ref{Fig:OPV2V_SOTA} and~\ref{Fig:DAIRV2X_SOTA} compare the proposed \texttt{CodeFilling} with previous methods in terms of the trade-off between detection performance and communication bandwidth for DAIR-V2X and OPV2VH+ datasets under homogeneous and heterogeneous settings, respectively. Detailed values can also be found in the appendix. Baselines include no collaboration ($\mathcal{O}_i$), Where2comm~\cite{HuWhere2comm:NeurIPS22}, HMViT~\cite{XiangHMViT:ICCV23}, V2VNet~\cite{WangV2vnet:ECCV20}, DiscoNet~\cite{LiLearning:NeurIPS21}, V2X-ViT~\cite{XuV2XViT:ECCV22}, AttFuse~\cite{XuOPV2V:ICRA22} and late fusion, where agents exchange the detected 3D boxes directly. Note that HMViT is specifically designed for heterogeneous settings by using domain adaption, while~\texttt{CodeFilling} is naturally compatible with heterogeneous settings without additional cost.
We see that \texttt{CodeFilling}: i) achieves a far-more superior perception-communication trade-off across all the communication bandwidth choices and various collaborative perception settings, including camera-only, lidar-only, and heterogeneous 3D detection;  ii) significantly improves the detection performance, especially under extremely limited communication bandwidth, improves the SOTA performance by 11.093/5.271/38.357\%, 14.75/28.516/28.372\% for LiDAR/camera/heterogeneous on DAIR-V2X and OPV2VH+ even when the bandwidth is constrained by a factor of 100K; and iii) outperforms previous communication-efficient SOTA, Where2comm, with significantly reduced communication cost: 1333/115/863, 1206/1078/252 times less on DAIR-V2X and OPV2VH+. 

Furthermore, for inference speed, \texttt{CodeFilling} (36/99ms) is comparable to Where2comm (34/94ms), and significantly faster than HMViT (90/1266ms) on DAIR-V2X/OPV2VH+. This communication efficiency ensures that agents are able to actively collaborate with each other. 

\noindent\textbf{Robustness to pose error and communication latency.}
We validate the robustness against pose error and communication latency on both OPV2VH+ and DAIR-V2X.  The pose error setting follows CoAlign~\cite{LuRobust:ICRA23} using Gaussian noise with a mean of 0m and standard deviations ranging from 0m to 1.0m. The latency setting follows SyncNet~\cite{LeiLatency:ECCV22}, varying from 0ms to 500ms.
Figs.~\ref{Fig:Robust_poseerror} and~\ref{Fig:Robust_latency} show the detection performances as a function of pose error and latency, respectively. We see: i) while perception performance generally declines with increasing levels of pose error and latency, \texttt{CodeFilling} consistently outperforms baselines under all imperfect conditions; ii) \texttt{CodeFilling} consistently surpasses No Collaboration, whereas baselines fail when pose error exceeds 0.6m and latency surpasses 100ms. 

\begin{figure*}[!ht]
  \centering

  \begin{subfigure}{0.19\linewidth}
    \includegraphics[width=1.0\linewidth]{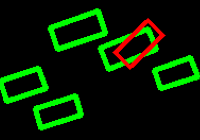}
    \caption{Ego detections}
    \label{Fig:vis_ego_input}
  \end{subfigure}
  \begin{subfigure}{0.19\linewidth}
    \includegraphics[width=1.0\linewidth]{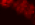}
    \caption{Ego feature}
    \label{Fig:vis_ego_feature}
  \end{subfigure}
  \begin{subfigure}{0.19\linewidth}
    \includegraphics[width=1.0\linewidth]{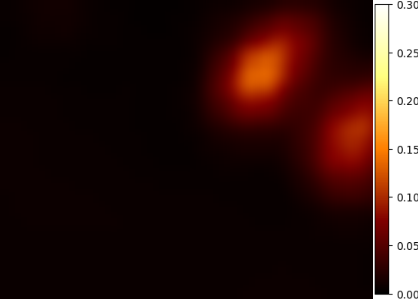}
    \caption{Ego information score}
    \label{Fig:vis_ego_conf}
  \end{subfigure}
  \begin{subfigure}{0.19\linewidth}
    \includegraphics[width=1.0\linewidth]{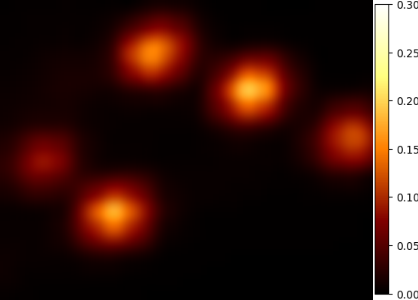}
    \caption{Collaborator score 1}
    \label{Fig:vis_C1_conf}
  \end{subfigure}
  \begin{subfigure}{0.19\linewidth}
    \includegraphics[width=1.0\linewidth]{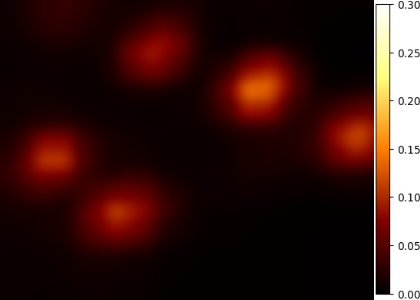}
    \caption{Collaborator score 2}
    \label{Fig:vis_C2_conf}
  \end{subfigure}

  \begin{subfigure}{0.19\linewidth}
    \includegraphics[width=1.0\linewidth]{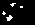}
    \caption{Selection matrix 1}
    \label{Fig:vis_m1}
  \end{subfigure}
  \begin{subfigure}{0.19\linewidth}
    \includegraphics[width=1.0\linewidth]{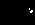}
    \caption{Selection matrix 2}
    \label{Fig:vis_m2}
  \end{subfigure}
  \begin{subfigure}{0.19\linewidth}
    \includegraphics[width=1.0\linewidth]{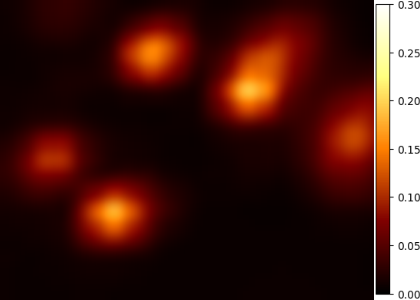}
    \caption{Filled confidence}
  \end{subfigure}
    \begin{subfigure}{0.19\linewidth}
    \includegraphics[width=1.0\linewidth]{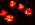}
    \caption{Codebook-based message}
    \label{Fig:codefilled}
  \end{subfigure}
  \begin{subfigure}{0.19\linewidth}
    \includegraphics[width=1.0\linewidth]{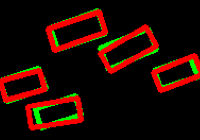}
    \caption{Collaborative detections}
    \label{Fig:final_result}
  \end{subfigure}

  \vspace{-3mm}
  \caption{Visualization of collaboration in~\texttt{CodeFilling}. \textcolor{green}{Green} and \textcolor{red}{red} denote ground truth and detection, respectively.}
  \label{Fig:Collaboration}
  \vspace{-4mm}
\end{figure*}

\begin{figure*}
\centering
\begin{subfigure}{0.19\linewidth}
    \includegraphics[width=1.0\linewidth]{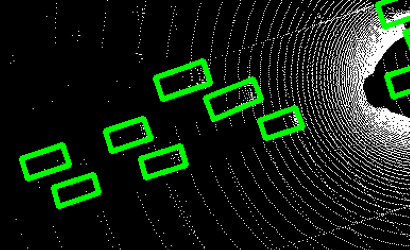}
    \caption{No Collaboration}
  \end{subfigure}
  \begin{subfigure}{0.19\linewidth}
    \includegraphics[width=1.0\linewidth]{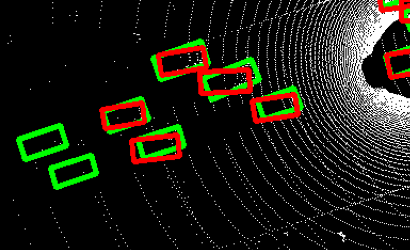}
    \caption{V2X-ViT}
  \end{subfigure}
  \begin{subfigure}{0.19\linewidth}
    \includegraphics[width=1.0\linewidth]{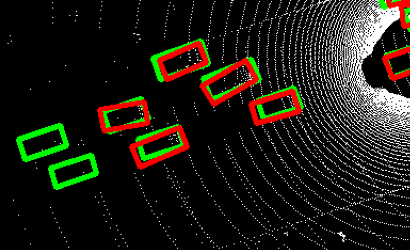}
    \caption{HMViT}
  \end{subfigure}
    \begin{subfigure}{0.19\linewidth}
    \includegraphics[width=1.0\linewidth]{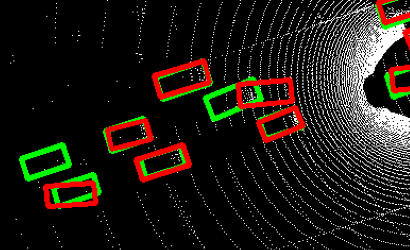}
    \caption{Where2comm}
  \end{subfigure}
  \begin{subfigure}{0.19\linewidth}
    \includegraphics[width=1.0\linewidth]{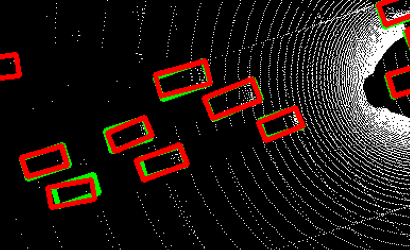}
    \caption{CodeFilling}
  \end{subfigure}
  \vspace{-4mm}
  \caption{\texttt{CodeFilling} achieves more accurate detections with 1256 times less communication cost. \textcolor{green}{Green} and \textcolor{red}{red} boxes denote ground-truth and detection, respectively.}
  \label{Fig:Vis_Comparison}
  \vspace{-8mm}
\end{figure*}

\vspace{-2mm}
\subsection{Ablation studies}\label{subsec:Ablation}
\vspace{-2mm}

\textbf{Effectiveness of our message selection and representation.} Fig.~\ref{Fig:Abl_comp} compares \texttt{CodeFilling}, the one without message selection, and the one without codebook and message selection.
We see that: i) applying code index representation reduces the communication cost by 208 times while maintaining the same detection performance, as it bypasses the high channel dimension typical of feature vectors, and an integer index requires less data than the original floating-point numbers, and ii) applying information-filling-driven message selection achieves 4.8\% higher detection performance with the same communication cost, as it reallocates the bandwidth wasted in redundant information to more beneficial information.

\noindent\textbf{Ablation of information-filling-driven message selection.}
Fig.~\ref{Fig:Abl_util} compares different utility designs in information-filling optimization: sum, max, and scenarios without selection. The max utility design favors selecting collaborators with the highest score, leading to no selection if the ego agent has the highest score.
We see that the sum-based utility outperforms both the max and no selection in the perception-communication trade-off across all communication bandwidth conditions. This superiority is due to the optimized combination of information from different collaborators, which has proved to be more effective than relying solely on the best-performing single agent. This approach encourages agents to participate in collaboration; even the top-performing agents can benefit from collaboration.

Fig.~\ref{Fig:Abl_util} evaluates three different information demands $u$ (0.5, 1.0, 1.5). We see that:
i) lower $u$ demonstrates better efficiency under limited communication budgets;
ii) higher $u$ demonstrates superior performance under ample communication budgets.
By adjusting $u$, \texttt{CodeFilling} consistently maintains superior performance-communication trade-off across all communication conditions.

\noindent\textbf{Ablation of codebook-based message representation.}
Fig.~\ref{Fig:Abl_size} and Fig.~\ref{Fig:Abl_residual} explore different codebook configurations: codebook size and code quantity. We see that:
i) all the codebook configurations demonstrate a superior perception-communication trade-off under highly constrained communication conditions, showing the effectiveness and robustness of the codebook-based representation;
and ii) larger codebook sizes and quantities yield better performance, while smaller sizes and quantities offer greater communication efficiency. By adapting the configuration, \texttt{ CodeFilling} maintains superior performance-communication trade-off across all communication budgets.



\vspace{-2mm}
\subsection{Qualitative evaluation}\label{subsec:Qualitative}
\vspace{-1mm}
\textbf{Visualization of message selection and representation.}
Fig.~\ref{Fig:Collaboration} showcases the efficient collaboration in \texttt{CodeFilling}. The scene features one ego agent and two collaborators - one with LiDAR and the other with a camera, with all visualizations presented from the ego's perspective.
Fig.~\ref{Fig:Collaboration} (a) and (j) compare the detection results before and after collaboration. We see that through collaboration, the ego agent successfully uncovers detections missed in its individual view.
Fig.~\ref{Fig:Collaboration} (d-g) displays the collaborators' information score maps and their corresponding selection matrices. We see that the redundant information in the overlapped regions is avoided, promoting communication efficiency.
Fig.~\ref{Fig:Collaboration} (c) and (h) illustrate the evolution of the information score before and after filling, highlighting that the information demands have been met, which is indicative of improved perception performance.
Fig.~\ref{Fig:Collaboration} (f) presents the codebook-represented message, revealing that: i) it is spatially sparse, and ii) the selected information from heterogeneous collaborators is unified in a common format, facilitating extensibility.

\noindent\textbf{Visualization of detection results.}
Fig.~\ref{Fig:Vis_Comparison} compares \texttt{CodeFilling} with previous SOTAs. We see that~\texttt{CodeFilling} qualitatively outperforms previous SOTAs with 1256 times less communication cost. The reason is that~\texttt{CodeFilling} avoids redundant information from different collaborators and employs efficient code index representation, thereby transmitting more critical perceptual information even with less communication cost.


\vspace{-2mm}
\section{Conclusions}\label{sec:conclusion}
\vspace{-2mm}

We propose \texttt{CodeFilling}, a novel communication-efficient collaborative 3D detection system with two novel designs: codebook-based message representation and information-filling-driven message selection. Extensive experiments covering both real-world and simulation scenarios show that \texttt{CodeFilling} not only achieves state-of-the-art perception-communication trade-off under various modalities, including LiDAR, camera, and heterogeneous settings but also is robust to pose error and latency issues.

\noindent
\textbf{Limitation and future work.} We plan to explore the temporal dimension and determine critical time stamps. 

\noindent\textbf{Acknowledgments.} This research is supported by the National Key R\&D Program of China under Grant 2021ZD0112801, NSFC under Grant 62171276 and the Science and Technology Commission of Shanghai Municipal under Grant 21511100900 and 22DZ2229005.


\clearpage
{
    \small
    \bibliographystyle{ieeenat_fullname}
    \bibliography{main}
}



\end{document}